%% file: main.tex
\documentclass[10pt, conference]{IEEEtran}

\input{macro}
\usepackage{url}
\usepackage{amsmath,amssymb}
\usepackage{enumitem}
\usepackage{pbox}
\usepackage{hyperref}
\usepackage{multirow}
\usepackage{xspace}
\usepackage[normalem]{ulem}
\usepackage[switch]{lineno}

\makeatletter 
\newcommand{\linebreakand}{%
  \end{@IEEEauthorhalign}
  \hfill\mbox{}\par
  \mbox{}\hfill\begin{@IEEEauthorhalign}
}
\makeatother 

\begin{document}

\title{Automated Repair of Programs from Large Language Models}

\author{
\IEEEauthorblockN{Zhiyu Fan}
\IEEEauthorblockA{
\textit{National University of Singapore}\\
Singapore \\
zhiyufan@comp.nus.edu.sg}
\and
\IEEEauthorblockN{Xiang Gao}
\IEEEauthorblockA{
\textit{Beihang University}\\
Beijing, China \\
xiang\_gao@buaa.edu.cn}
\and
\IEEEauthorblockN{Martin Mirchev}
\IEEEauthorblockA{
\textit{National University of Singapore}\\
Singapore \\
mmirchev@comp.nus.edu.sg}

\linebreakand 

\IEEEauthorblockN{Abhik Roychoudhury}
\IEEEauthorblockA{
\textit{National University of Singapore}\\
Singapore \\
abhik@comp.nus.edu.sg}
\and
\IEEEauthorblockN{Shin Hwei Tan}
\IEEEauthorblockA{
\textit{Southern University of Science and Technology}\\
Shenzhen, China \\
tansh3@sustech.edu.cn}
}

\maketitle

\begin{abstract}
Large language models such as Codex, have shown the capability to produce code for many programming tasks. However, the success rate of existing models is low, especially for complex programming tasks. One of the reasons is that language models lack awareness of program semantics, resulting in incorrect programs, or even programs which do not compile. In this paper, we systematically study whether automated program repair (APR) techniques can fix the incorrect solutions produced by language models in \leet{} contests.
The goal is to study whether APR techniques can enhance reliability in the code produced by large language models. Our study revealed that: 
(1) automatically generated code shares common programming mistakes with human-crafted solutions, indicating APR techniques may have potential to fix auto-generated code; 
(2) given bug location information provided by a statistical fault localization approach, the newly released Codex edit mode, which supports editing code,  is similar to or better than existing Java repair tools TBar and Recoder in fixing incorrect solutions. By analyzing the experimental results generated by these tools, we provide several suggestions: (1) enhancing APR tools to surpass limitations in patch space ({\em e.g.}, introducing more flexible fault localization) is desirable; (2) as large language models can derive more fix patterns by training on more data, future APR tools could shift focus from adding more fix patterns to  synthesis/semantics based approaches, (3) combination of language models with APR to curate patch ingredients, is worth studying.
\end{abstract}

\input{introduction}

\input{setting}

\input{rq1}
\input{rq2}
\input{rq3}

\input{rq5}
\input{rq4}
\input{threats}
\input{related}
\input{conclusion}

\bibliographystyle{IEEEtran}
\bibliography{references}

\end{document}


\title{Supplementary: Automated Repair of Auto-generated Codex Programs}

\maketitle

\input{exapme-table2}

%% file: macro.tex
\newcommand{\leet}{LeetCode}

\newcommand{\numtasks}[0]{113\xspace}

\usepackage{listings}
\usepackage{xcolor}
\definecolor{main-color}{rgb}{0.6627, 0.7176, 0.7764}
\definecolor{string-color}{rgb}{0.3333, 0.5254, 0.345}
\definecolor{key-color}{rgb}{0.8, 0.47, 0.196}
\lstdefinestyle{mystyle}
{
    language = Java,
    basicstyle = {\ttfamily \color{main-color}},
    stringstyle = {\color{string-color}},
    keywordstyle = {\color{key-color}},
    keywordstyle = [2]{\color{lime}},
    keywordstyle = [3]{\color{yellow}},
    keywordstyle = [4]{\color{teal}},
    morekeywords = [3]{<<, >>},
    morekeywords = [4]{++},
    basicstyle=\ttfamily\scriptsize,
    commentstyle=\color{blue}\ttfamily,
    morecomment=[f][\lstbg{red!20}]-,
    morecomment=[f][\lstbg{green!20}]+,
    morecomment=[f][\lstbg{yellow!20}]++,
    morecomment=[f][\lstbg{yellow!20}]--,
    morecomment=[f][\textit]{@@},
    texcl=false
}

\usepackage{color, colortbl}
\definecolor{grey}{rgb}{0.7,0.7,0.7}

\newcommand{\lstbg}[3][0pt]{{\fboxsep#1\colorbox{#2}{\strut #3}}}
\lstdefinelanguage{diff}{
  basicstyle=\ttfamily\scriptsize,,
  morecomment=[f][\lstbg{red!20}]-,
  morecomment=[f][\lstbg{green!20}]+,
  morecomment=[f][\lstbg{yellow!20}]++,
  morecomment=[f][\textit]{@@},
  texcl=false
}

\usepackage{tcolorbox}

\newcommand{\totalBeforePatch}{46\xspace}

\newcommand{\numOfUnsolvedSolution}{335\xspace}
\newcommand{\numOfUnsolvedTask}{67\xspace} 
\newcommand{\numOfSolvedTask}{46\xspace} 
\newcommand{\algoin}{191}
\newcommand{\mydataset}{LMDefects}
\newcommand{\antipattern}{anti-patterns}

\usepackage{tikz}
\newcommand*\circled[1]{\tikz[baseline=(char.base)]{ \node[shape=circle,draw,inner sep=0.6pt] (char) {#1};}}

%% file: introduction.tex
\section{Introduction}
\label{sec:intro}

Designing AI-based systems to automatically solve programming tasks has gained considerable attention in recent years.
The most notable of these comes in the form of transformer-based large-scale language models, which can be used to achieve impressive performance in generating text.
The transformer-based models, such as Codex~\cite{codex} and AlphaCode~\cite{alphacode}, have successfully generated code for many programming tasks in Python, Java, and C.
Technically, these techniques treat code generation as a transformation problem, which takes as input natural language descriptions and transforms them into programming language.

Although transformer-based models successfully solved many programming tasks, their success rate is still relatively low.
When evaluating on \emph{pass@5} metric~\cite{codex}, the best Codex model achieves 24.52\% passing rate at introductory-level tasks and 3.08\% passing rate at competition-level tasks~\cite{codex} from APPS dataset~\cite{apps}.
The best AlphaCode model achieves 20.36\% and 7.75\% passing rates on introductory-level and competition-level tasks, respectively~\cite{alphacode}.
Lacking deep understanding of task descriptions and program semantics are the main reasons that cause the low success rate.
Transformer-based models treat code generation as a sequence-to-sequence transformation by treating description and code as token sequences which cannot capture deep semantic features of programs.
In contrast, generating entire programs requires an understanding of the entire task's description which usually comprises complex logic, and figuring out the solutions to programming tasks relies on deep algorithm reasoning. Although it is important to systematically study the reasons behind the ineffectiveness of language models in solving programming tasks, there is little to no study that characterizes the defects made in the programs automatically generated by language models, creating a gap in understanding how to further improve these automatically generated programs. 

\input{table/keyfindings}

Automated program repair (APR) is an emerging area for automated rectification of programming errors~\cite{apr}.
APR techniques take as inputs a buggy program and a correctness specification, and produce a fixed program by slightly changing the program to make it satisfy the given specification.
Typical repair tools generate patches by reasoning about the program semantics against the given specification.
For instance, semantic-based repair tools (e.g., SemFix~\cite{semfix}, Angelix~\cite{angelix}) generate patches via symbolic execution, while search-based repair tools (e.g., GenProg~\cite{genprog}, TBar~\cite{tbar}) search for correct patches among a pre-defined search space. 
APR has shown promising results in fixing real-world bugs 
but they are still limited to generating small patches (usually one-line fixes) due to the complexity of semantic reasoning, and search space explosion -- when considering multi-line fixes.

The strength and weakness of language models and APR techniques inspire us to think about the following question: \textit{can automated program repair improve the code produced by language models?} In this paper, we apply existing APR techniques to the code generated by the Codex model, and answer the following research questions:

\smallskip

\noindent
\textit{\textbf{(RQ1) What mistakes are common in  auto-generated code?}}

We study the bug patterns of code produced by Codex, and whether they are similar to bugs in human-written code. 

\smallskip
\noindent
\textit{\textbf{(RQ2) Can APR tools effectively fix code from Codex?}}

We study how effective APR tools (TBar and Recoder) are in fixing the code produced by Codex. 

\smallskip
\noindent
\textit{\textbf{(RQ3) Can Codex edit mode fix program bugs?}}

In March 2022, a new version of Codex was released~\cite{codex-e}, which can edit existing content in a complete program rather than just completing a partial program.
Codex edit mode (we call this mode \textit{Codex-e}  throughout this paper) requires users to provide instructions to guide the revision, such as ``translate the java program to javascript'' \cite{codex-e}. To fix a bug, users need to provide precise and clear instructions. How to automatically produce such instructions still remains an open question. We study whether the side effect of APR tools, such as fault localization results, can be used to guide Codex-e, and how effective Codex-e is in fixing program bugs.

Table~\ref{tab:findings} presents the key findings of our study.
Our result shows that existing APR tools (pattern-based and learning-based APR) are still quite limited, including limited patch space, fix locations and patch size --- thus enhancing APR tools to surpass these limitations (e.g., introducing a more flexible fault localization strategy) is highly desirable. Specifically we see possible collaboration between APR tools and Codex-e for curating patch ingredients to construct complex patches.


\noindent
\paragraph*{ Contributions}
The contributions of this paper are:
\begin{itemize}[leftmargin=*, topsep=0pt]
\item We present a systematic study of automated repair of buggy programs produced from language models.
\item  To the best of our knowledge, we conduct the first study that evaluates the efficacy of the newly released Codex edit mode as an automated repair tool.
\item We propose \mydataset{}, a new dataset that contains 113 Java programming tasks. Among them, \totalBeforePatch tasks have been successfully solved by Codex and \numOfUnsolvedTask{} of them remain unsolved. 
\end{itemize} 




%% file: table/keyfindings.tex
\begin{table*}[!t]
\caption{Our key finding and implications on the bug patterns made by Codex and the effectiveness when applying existing repair tools and Codex-e to fix these bugs.}
\label{tab:findings}
\begin{tabular}{|p{0.47\linewidth} | p{0.48\linewidth}|}\hline
\rowcolor{grey}
Findings on Bug Pattern (Section~\ref{sec:rq1})  & Implications \\\hline
Auto-generated code share common mistakes with human programmers. 57\% of bugs made by Codex are algorithm-related, and 11\% of them are due to syntax errors. To fix the remaining bugs made by Codex. 13.4\% of them require small changes (e.g., changing operator and replacing variables), 18.5\% of them require larger patches. &  As Codex generated code share common mistakes with human-written code, using APR techniques to enhance reliability in auto-generated code by automatically fixing the bugs in the auto-generated code is worth studying. 
\\\hline
Auto-generated code contain negative symptoms or undesirable code patterns such as: \circled{1} names that indicate wrong algorithms; \circled{2} similar code blocks (code smells related); \circled{3} producing irrelevant helper functions.  & Instead of depending on token log-probability, language model designers can consider incorporating rigorous code quality checks from the perspective of a program itself to enhance code generation and recommendation. 
\\\hline\hline

\rowcolor{grey}
Findings on APR's effectiveness (Section~\ref{sec:rq2}) & Implications \\\hline
\multirow{2}{\linewidth}{Existing pattern-based and learning-based APR approaches can fix a small number of bugs in auto-generated code. The challenges in fixing auto-generated code include: \circled{1} limited search space; \circled{2} unable to generate multi-edit patches; \circled{3} lack of awareness of program dependencies.}& 
Manually designing fix patterns is not scalable, and future research may either need to look more at program synthesis based approaches, or need to curate patterns automatically from huge training data. \\\cline{2-2} 
 & 
 Statistical fault localization is widely used by APR tools to determine fix location, which might be limited. Advanced fix localization techniques based on program dependency analysis may help to improve APR's repairability.
\\\hline\hline
\rowcolor{grey}
Findings on Codex Edit Mode (Codex-e) (Section~\ref{sec:rq3})            & Implications \\\hline
Given ``proper" instructions (such as where to fix), Codex-e even outperformed pattern-based and learning-based APR tools. With/Without controlling the fault locations affect the characteristics of generated patches by Codex-e. So, ``what kind of guidance should be given to Codex-e?'', needs to be further studied.
& 
 Considering the similar effectiveness of Codex-e with and without location guidance, future APR research should strike a balance between controlling the fault location and providing flexibility in the fault location (allowing it to generate multi-hunk patches). This work would be along the lines of  engineering prompts for language model based code generators.  
\\\hline
Codex-e is able to generate patches at flexible locations beyond the given location or statement. This enables Codex-e to produce more correct and larger patches, especially when the given location is not precise. & 
Future APR tools could explore more flexible forms of fix localization to allow fixes to be generated at multiple locations. \\\hline
\hline

\rowcolor{grey}
Findings on Combining Search Space of different Tools (Section~\ref{sec:rq3})  & Implications \\\hline
Combining the search space of different tools (TBar and Codex-e) could produce the required patch ingredients to fix more incorrect solutions.
& Combination of APR tools with language model based tools like Codex-e, for curating patch ingredients --- is worth studying. One can combine multiple incorrect solutions produced by Codex to get more patch ingredients. APR tools can consider using Codex as a source of crafting rich patch ingredients. 
\\\hline

\end{tabular}
\end{table*}

%% file: setting.tex
\begin{figure*}[!t]
    \centering
    \includegraphics[width=0.85\textwidth]{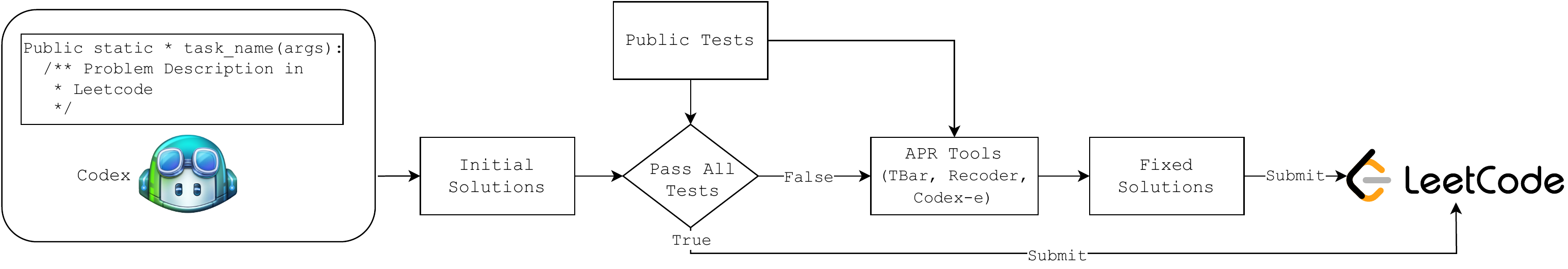}
    \vspace{-3pt}
    \caption{The workflow of automatically fixing programs generated by Codex}
    \label{fig:workflow}
    \vspace{-6pt}
\end{figure*}

\section{Study Setting}
In this section, we present the setting of our study, including the overall workflow, the Codex model, parameters, dataset, APR tools, etc. All experiments were conducted on an Ubuntu-16.04 server, with 64GB RAM and Intel(R) Xeon(R) CPU E5-2660 @ 2.00GHz, and NVIDIA Titan V GPU.

\paragraph{\textbf{Codex Model}}
Codex~\cite{codex} is the model that powers GitHub Copilot \cite{copilot}
 which completes a program given a natural language prompt.
Codex supports many programming languages (e.g., Python, C/C++, Java).
Their training data contains both natural language and billions of lines of public code from GitHub.
In our study, we use the pre-trained Codex code-davinci-002 and Codex-e code-davinci-edit-001 model~\cite{codexmodel}, which were both trained on data up to Jun 2021.

\paragraph{\textbf{Methodology and Dataset}}
Figure~\ref{fig:workflow} shows the overall workflow of our study. 
We first use Codex to generate initial solutions for each task and validate the correctness of generated solutions on {\em public example test cases}.
For each unsolved programming task, existing test-based repair tools (using the public tests) are then applied to fix the incorrect solutions produced by Codex. The patched solutions are then validated using (1) the public tests, and (2) the held-out (private) tests in the \leet{} platform.
To answer our research questions, we build a dataset \mydataset{} with \numtasks programming tasks in \leet~\cite{leetcode}. 
\leet{} is an online judge platform, 
which includes over 2,300 different problems, ranging from easy to hard level.
It also has a forum \cite{Leetforum} with active community where correct solutions for each programming task can be found (important for our manual analysis of incorrect solutions).
Each task is described using natural language text accompanied with {\em  1--3 public tests} that provide examples with pairs of (input, output). When a solution is submitted to \leet{}, it runs a set of private tests to validate the correctness of submissions. \leet{} has weekly and biweekly contests, where it releases new programming tasks.
In our study, we only consider easy-level and medium-level problems because Codex fails to solve most hard problems~\cite{codex} (we also exclude seven tasks that require customized data structures that Codex is unlikely to handle).
To prevent the case where the collected dataset was used in the training set of Codex, we only consider contests that are released after Jun 2021 (the end date where the Codex training data is extracted from). 
Overall, we crawl through all contests in LeetCode from 4 July 2021 until 6 Apr 2022. This leads to a total of 40 weekly contests and 20 biweekly contests. In total, \mydataset{} contains 60 easy and 53 medium-level programming tasks. 
Several datasets with programming tasks exist~\cite{codex,alphacode,apps,tan2017codeflaws,Caballero_Description2Code_Dataset_2016,codenet}. They are either based on contests from programming competition platforms (e.g., Codeforces) or hand-written programming tasks. We do not use existing datasets because (1) Codex was already trained on GitHub where solutions for many previous programming tasks exist (e.g., APPS, CodeContest) 
(2) some programming tasks do not have public tests which is a prerequisite for APR techniques (e.g., HumanEval~\cite{codex}), (3) most APR tools only support Java programs, whereas the HumanEval dataset curates Python programs.

\paragraph{\textbf{Prompt and Parameters}}
Codex model takes as inputs a \emph{prompt}, which is the combination of natural language text and code snippet, where the natural language text represents the programming task description and the code snippet is the starting point for language model to complete the code. We evaluate Codex in zero-shot prompt settings (does not include example input/output in the prompt). Note that public test cases can also be embedded in the prompt (namely \emph{n-shots prompt}) 
but we feed the public tests in the natural language description to guide APR tools instead of embedding them in the prompt. 
Lines 1--3 in Figure~\ref{listing:example-dp} presents an example prompt that we use for a \leet{} programming task~\footnote{https://leetcode.com/problems/minimum-operations-to-convert-number/}. 
Given such a prompt, we run Codex to generate 50 candidate solutions and select the top five solutions with the highest probability of being correct (via the \emph{best\_of} parameter of Codex). The selected solutions are first validated by running against the public tests (e.g., lines 14--17 in Figure~\ref{listing:example-dp}, we manually convert the public tests into JUnit tests), and then validated against private tests by submitting them to \leet{}.
We reuse the same setting as prior work~\cite{codex}. Specifically, we (1) set temperature to 0.8 (which has the best performance when generating 50 candidate solutions), and (2) prepare the stop sequences to be ``public'', ``class'', ``//'', ``System.out.print''. We set the ``max token length'' to 2048.


\begin{figure}[!t]
\centering
\begin{lstlisting}[style = mystyle, escapechar=@]
 public static int minimumOperations(int[] nums) {
/** You are given a 0-indexed array nums consisting of n positive integers. ...
 * Return the minimum number of operations required to make the array alternating.*/
  int n = nums.length;
  int[] dp = new int[n];
  dp[0] = 0; dp[1] = 1;
  for (int i = 2; i < n; i++){
    dp[i] = dp[i - 1] + 1;
    if(nums[i] == nums[i - 2])
      dp[i] = Math.min(dp[i - 2] + 1, dp[i]);
  }
  return dp[n - 1];
}
// Public Test 1:
// Input: nums = [3,1,3,2,4,3], Output: 3
// Public Test 2:
// Input: nums = [1,2,2,2,2], Output: 2
\end{lstlisting}
\caption{An example of misaligned algorithm, taken from LeetCode programming task \textit{minimumOperations} on weekly-contest-280.} 
\label{listing:example-dp}
\vspace{-6pt}
\end{figure}


\paragraph{\textbf{APR Tools}}
To evaluate whether repair tools can fix the incorrect solutions produced by Codex, we evaluate on two Java APR tools 
because Java APR tools have been widely studied, and many of them are open-source. Among all the open-source Java APR tools, we select TBar and Recoder because (1) they are the most recent representative of different approaches (i.e., TBar represents a search-based and pattern-based APR tool, whereas Recoder is a learning-based approach), and (2) these tools have reported the best results by generating the highest number of correct patches on the Defects4J~\cite{defects4j} benchmark (almost all Java APR tools have been evaluated on it). 
Since both TBar and Recoder are test-driven APR tools, we use the public test cases given in the program description to guide the repair, while the private test cases are applied to validate the patched solutions. 
We run TBar and Recoder in default settings, and the repair process stops if a patch that passes all public tests is found. 
We set the timeout to 15 minutes, following the time limit used in prior work on automated repair of programming assignments~\cite{yi2017feasibility}.
As Codex edit mode (Codex-e) can modify existing code by generating program edits, we investigate whether Codex-e can serve as an APR tool and compare it  with TBar and Recoder. 


%% file: rq1.tex
\section{RQ1: What mistakes are common in auto-generated code?}
\label{sec:rq1}

\begin{table*}[!t]
\setlength{\tabcolsep}{2.8pt}
\setlength{\extrarowheight}{5pt}
\caption{Defect Classification of incorrect solutions}
\vspace{-6pt}
\label{tab:bug-pattern}
\footnotesize
\centering
    \begin{tabular}{l|l|l|r|r|r}\hline
        Defect Category & Sub-category & Definition & Easy & Medium & Total\\\hline
\hline
\multirow{3}{*}{ \textbf{M}ulti-hunk}
        & (M-S) \textbf{S}imilar  & 
\pbox{8cm}{Similar single-hunk bugs (require similar fixes) exist at multiple discontinuous program locations}
 & 7 & 2 & 9\\\cline{2-6} 
        
        & (M-U) \textbf{U}nique & 
        \pbox{8cm}{Distinct single-hunk bugs exist at multiple discontinuous program locations, and the total lines of patches are no more than five lines}
& 19 & 20  & 39\\\cline{2-6} 
        & (M-L) Need \textbf{L}arge Fix & 
        \pbox{8cm}{The bug is (1) neither M-S or M-U and (2) needs to edit more than five lines at multiple locations} 
& 5 & 9 &  14\\\hline
\multirow{8}{*}{\textbf{S}ingle-hunk~\cite{hercules}}
        & (S-O) \textbf{O}perator Mutation  & 
\pbox{8cm}{Replace arithmetic/logical/relational/bitwise operator with another operator or insert/delete operators and relevant operands or modify operator precedence}
        & 7 & 5  & 12  \\\cline{2-6} 
& (S-C) \textbf{C}onstant Mutation & 
\pbox{8cm}{Replace constant (not in array or function call) with a variable/constant/function call}
        & 3 & 0  & 3   \\\cline{2-6} 
        & (S-V) \textbf{V}ariable Mutation &
\pbox{8cm}{Replace variable (not in array or function call) with a variable/constant/function call}
        & 2 & 0  & 2  \\\cline{2-6} 
& (S-A) \textbf{A}rray Mutation & 
\pbox{8cm}{Replace the array access with other constant/variable, operands with arithmetic operators, or replace an array with another array}   
        & 1 & 0  & 1   \\\cline{2-6}
& (S-F) \textbf{F}unction Call Mutation & 
\pbox{8cm}{Replace function call with another function call or change function arguments}
 & 2 & 1  & 3   \\\cline{2-6}
        &(S-AS) \textbf{A}dd \textbf{S}tatements & 
\pbox{8cm}{Insert a continuous chunk of statements}
        & 8 & 1  & 9  \\\cline{2-6} 
        &(S-DS) \textbf{D}elete \textbf{S}tatements & 
\pbox{8cm}{Delete a continuous chunk of statements}     
        & 2 & 3  & 5 \\\cline{2-6}
        &(S-HO) \textbf{H}igher \textbf{O}rder  & 
\pbox{8cm}{A single-hunk patch that combines multiple single-hunk bugs}
        & 5 & 5  & 10 \\\cline{2-6}
\hline
Algorithm-related
        & Misaligned Algorithm & \pbox{8cm}{The algorithm used is misaligned with the requirement given in the task description}  & 30 & 161 & \algoin{} \\\hline
\multirow{2}{*}{Syntax Error}
& Incomplete Code  & \pbox{8cm}{For the last line of the program, only parts of the program is printed} & 7 & 17 & 24 \\\cline{2-6}
& Invoke Undefined Program Elements & \pbox{8cm}{Fails to compile due to invoking undefined variables/functions/classes} & 2 & 11 & 13 \\\cline{2-6}
\rowcolor{grey}   
    Total   & - & - & 100 & 235 & \numOfUnsolvedSolution\\
        \hline
\end{tabular}
\vspace{-6pt}
\end{table*}


Before we apply APR techniques in fixing the automatically generated solutions, we investigate its feasibility by analyzing the typical mistakes made in solutions produced by Codex.

Given a programming task in \mydataset{} for Codex to solve, we first run the five auto-generated solutions on the public tests 
and submit them to \leet~online judge platform to validate using private tests. If an auto-generated solution $s$ by Codex fails to pass all public and private tests, we consider $s$ an \emph{incorrect solution}.
If all the five auto-generated solutions for a programming task are incorrect solutions, we consider this task as \emph{unsolved}. 
Overall, \numOfSolvedTask programming tasks can be \emph{solved} by Codex.
We study the mistakes of \numOfUnsolvedSolution \emph{incorrect solutions} $S_{buggy}$ in the remaining \numOfUnsolvedTask \emph{unsolved} programming tasks that lead to compilation errors or test failures. 

For each incorrect solution $s_{buggy} \in S_{buggy}$, two annotators (two authors of the paper) separately and manually fix it by first referring to other solutions in \leet~discussion forum for repair hints, and then constructing a minimal patch that fixes the bugs. 
The constructed patch for each incorrect solution is cross-validated by the two annotators who make sure the patched solution $s_{fixed}$ is accepted by \leet~platform. 
Our goal is to construct a ``ground truth'' patch $s_{fixed}$ for each incorrect solution to obtain  the ``diff'' between $s_{buggy}$ and $s_{fixed}$. 
Based on this ``diff'', two authors manually classify each $s_{buggy}$ using the defect categories in Table~\ref{tab:bug-pattern}.
Each $s_{buggy}$ is assigned to one defect category. If there is any disagreement during the ground truth construction or defects classification, annotators discuss with other authors to resolve the disagreement (there were 14 initial disagreements, all of which were successfully resolved).

We derive the defect classification based on categories used in Codeflaws~\cite{tan2017codeflaws} (a benchmark that contains incorrect submissions by participants in programming competitions).
The detailed classification and their definitions are shown in Table~\ref{tab:bug-pattern}.
It also shows the number of incorrect solutions (both ``Easy'' and ``Medium'') belongs to each category.
The example code of each defect category can be found in the supplementary material.
The defect classification in auto-generated code overlap with those in Codeflaws. Specifically, both Codeflaws and our dataset contain defects where either multi-hunk or single-hunk fixes are required. Moreover, for the single-hunk fixes, both datasets share similar mutation operators (e.g., operator mutation, and variable mutation). This indicates Codex made similar programming mistakes as human participants. 
We think this is expected because Codex is trained with a lot of human-written programs that can be potentially buggy.
Besides the above single and multi-hunk bugs, syntax errors and algorithm-related errors are prevalent in Codex generated solutions. We manually analyzed these solutions to study the root causes behind these errors. 

\noindent \textbf{Syntax Errors.} Our manual analysis revealed that auto-generated programs that lead to compilation errors usually have (1) incomplete code, or (2) invoking undefined variables/functions/classes. To reduce the likelihood of Codex in generating incomplete code, we select the maximum token length allowed (i.e., 2048 tokens) by Codex for generating 50 candidate solutions. Despite providing the maximum length as the bound for code generation, Codex still generates incomplete code where the average token length is 628.  It is worthwhile to study the feasibility of applying code completion techniques for fixing the auto-generated incomplete code by Codex. Meanwhile, for programs with undefined functions, one needs to synthesize the function body to resolve the compilation errors. \emph{Future research can work on using program synthesis techniques to resolve the undefined functions or invoking Codex on a function-by-function basis to synthesize the function body}. Apart from these compilation errors, we also observe that Codex is prone to generate programs which fail to compile due to a missing/extra close bracket at the end of the program (in total, there are 23 of these cases). Since bracket mismatch can be fixed easily (using a regular expression matching mechanism), we manually fix them and further classify their defects into defect categories in Table~\ref{tab:bug-pattern}. 


\noindent \textbf{Misaligned Algorithm.} Among all incorrect solutions, \algoin{} solutions use wrong algorithms to solve the given tasks, including TLE (Time Limit Exceeded). The problem of generating solutions that do not meet the user intention is known as the \emph{misalignment} problem~\cite{codex}. All defects classified as ``misaligned algorithm'' suffer from the misalignment problem. 


\noindent \textbf{Negative Symptoms in Auto-generated code.} Auto-generated patches are known to exhibit certain \emph{\antipattern{}} (program transformations  leading to nonsensical patches)~\cite{tan2016anti}. Inspired by this, we analyze whether the code generated by Codex contains negative symptoms that are unlikely to be correct programs by studying the ``Misaligned Algorithm'' and ``Syntax Error'' categories: 

\begin{description}[leftmargin=*]
\item[\circled{1} \emph{Names indicate Wrong Algorithms:}]
In the ``Misaligned Algorithm'' category, Codex is prone to generate solutions with certain variable names which points to the underlying algorithm/data structure used is wrong. Figure~\ref{listing:example-dp} shows an example of using ``dp'' variable, where the algorithm used (i.e., ``dp'' refers to dynamic programming) is incorrect. We think Codex solves the task with dynamic programming because it is misled by other programs named ``minimumOperations'' but for a different programming task in GitHub. Similarly, we observe other variable names such as ``pq'' (priority queues), ``q'' (queue) that indicate the wrong data structures being used.

\item[\circled{2} \emph{Similar Code Blocks:}] We notice that when Codex struggles to find a high quality solution for a given prompt, it tends to repeatedly generate similar code blocks (code clones with minor variations in variable names, statement usages, and control structures). Figure~\ref{listing:example-similar-code} shows a solution with similar code blocks (only variable names differ) at lines 5--8 and 9--12.
\item[\circled{3} \emph{Irrelevant Helper Functions:}] Although we reuse Codex's setting in adding stop sequences (for terminating code generation once we reach the end of a function), we observe that Codex is still likely to produce redundant helper functions which are irrelevant to the given prompt.
\end{description}

\begin{figure}[!t]
\centering
\begin{lstlisting}[style = mystyle, escapechar=@]
public static int minimumSum(int num) {/** ...*/
 String str = "" + num;
 int first = Integer.MAX_VALUE;
 int second = Integer.MAX_VALUE; ...
++ if (firstNum.length() == 1) {
++  first=firstNum.charAt(0) - '0';
++ } else {
++  first=Integer.parseInt(firstNum.toString());}
++ if (secondNum.length() == 1) {
++  second=secondNum.charAt(0) - '0';
++  } else {
++  second=Integer.parseInt(secondNum.toString());}
...
\end{lstlisting}
\caption{An example of generating similar code block (highlighted with ``++''), taken from LeetCode programming task \textit{minimumSum}.}
\label{listing:example-similar-code}
\vspace{-10pt}
\end{figure}

\begin{tcolorbox}[left=2pt,right=2pt,top=2pt,bottom=2pt]
\vspace{-3pt}
Auto-generated programs share common mistakes with human-written programs, and contain certain negative symptoms including: (1) names indicate wrong algorithms; (2) similar code blocks; (3) irrelevant helper functions.
\vspace{-3pt}
\end{tcolorbox}

%% file: rq2.tex
\section{RQ2: How effective are APR tools in fixing the code produced by Codex?}
\label{sec:rq2}



Given the compilable incorrect solutions by the Codex model, we run TBar and Recoder to assess their ability in generating patches. During the patch validation stage, the automatically generated patches are categorized as below:

\begin{table}[!t]
\caption{The number of patches and fixed tasks produced by TBar and Recoder (include both single-hunk and multi-hunk)}
\vspace{-6pt}
\center
\label{tab:correct_vs_plausible}
    \begin{tabular}{l|rr|rr}
    \hline
    Tool & \multicolumn{2}{c|}{Correct/Plausible patches} & \multicolumn{2}{c}{Correctly Fixed Tasks}\\\cline{2-5}
     & easy & medium & easy & medium \\\hline
    TBar & 6/16 & 3/22 & 3 & 3\\\hline
    Recoder & 6/16 & 5/20 & 3 & 5\\\hline
\end{tabular}
\vspace{-6pt}
\end{table}

\begin{description}[leftmargin=*]
\item [Plausible patches.] Plausible patches are patches that make the incorrect solutions pass the given public tests.
\item[Correct patches.] Correct patches are patches that make the incorrect solutions pass both the public tests and private tests and accepted by \leet{}.


\end{description}

\begin{figure}[!t]
\centering
\begin{lstlisting}[style = mystyle, escapechar=@]
// task delete-characters-to-make-fancy-string 
public static String makeFancyString(String s){...
    if(...) {
      sb.deleteCharAt(i);
-      i -= 2;
+      i -= 1;@ \color{blue}\ttfamily// constant mutation (S-C-3)@
    }...}
// task watering-plants
public static int wateringPlants(int[] plants, int capacity) {...
  if (plants[i] > currWater) {
    steps += (i - 1) * 2;
+   steps++;@ \color{blue}\ttfamily//add a statement (S-O-10)@
    ...} ...}
\end{lstlisting}
\vspace{-6pt}
\caption{Two incorrect solutions fixed by Recoder but not TBar.}
\label{listing:tbar_vs_recoder}
\vspace{-6pt}
\end{figure}

Table~\ref{tab:correct_vs_plausible} shows the number of generated patches and the number of correctly fixed programming tasks by TBar and Recoder, respectively. 
Although TBar produces 16 and 22 plausible patches on easy-level and medium-level tasks, it only produces 6 easy and 3 medium correct patches.
Compared to TBar, Recoder produces less plausible patches (16 and 20 on easy and medium level, respectively), and more correct patches (6 and 5).
The ``Correctly Fixed Tasks'' columns of Table~\ref{tab:correct_vs_plausible} show the number of programming tasks correctly fixed by TBar and Recoder. Note that each programming task corresponds to the five selected incorrect solutions.
If any of these solutions is correctly fixed (accepted by \leet{}), we consider that this task has been solved. Overall, Recoder fixes eight programming tasks whereas TBar only fixes six tasks. Combining both tools, APR tools help Codex solve four more easy-level and five more medium-level tasks.

We further analyze the type of defects fixed by the two APR tools. Table~\ref{tab:data_apr} shows the number of solutions that can be correctly fixed for each defect category, where the ``TBar'' and the ``Recoder'' columns show the number of patches produced by the corresponding tools.
For each category, the repair tools may not fix the bug by minimally changing the program (i.e., repair tools may fix a bug using different operators than the minimal fix shown in the ``Defect sub-category'' column). The results show that existing APR tools are still limited in generating complex patches that require edits of multiple lines.

Figure~\ref{listing:tbar_vs_recoder} shows two examples where Recoder outperforms TBar. In the first example, despite having the ``Mutate Literal Expression'' pattern, TBar fails because it cannot find the correct literal to replace due to limited patch space. For the second example, TBar fails to generate the correct patch because it does not have the ``insert statement'' pattern.

For tasks that require multi-line fixes, both TBar and Recoder fail to generate any correct patches, one of the reasons is that the widely adapted statistical fault localization techniques in TBar and Recoder focus on identifying each faulty line separately, without considering program dependency among the suspicious lines. For example, to fix the bug in Figure~\ref{listing:minMove}, one needs to (1) change $s.length()-2$ to $s.length()$, and (2) simplify the if-condition. Using statistical fault localization, APR tools will generate patches for line 4 and lines 5--6 separately (without noticing that after fixing the if-condition at lines 5--6, the for-loop condition no longer need the extra ``-2'' at line 4 to prevent the ``IndexOutOfBoundException''). 


 \begin{figure}[!th]
\centering
\begin{lstlisting}[style=mystyle, escapechar=@]
public static int minimumMoves(String s) {
    //S-HO-5
    int count = 0; 
-   for (int i = 0; i < s.length() - 2; i++) {
-   if (s.charAt(i)==s.charAt(i+1) && s.charAt(i+1)
-      ==s.charAt(i + 2) && s.charAt(i)=='X') {
+   for (int i = 0; i < s.length(); i++) {
+   if (s.charAt(i)=='X') {
      count++;
      i += 2;
    }}
    return count;}
\end{lstlisting}
\vspace{-6pt}
\caption{An incorrect solution that should be fixed by modifying line 3 and lines 4--5 together.}
\label{listing:minMove}
\vspace{-6pt}
\end{figure}
    
\begin{table}[!t]
\centering
\caption{The number of correctly fixed solutions by TBar and Recoder, refer Table \ref{tab:bug-pattern} for abbreviation of defect classification}
\vspace{-6pt}
\setlength{\tabcolsep}{2pt}
    \begin{tabular}{l|rr|rr|rr}
         \hline
        Defect  & \multicolumn{2}{c|}{Total} & \multicolumn{2}{c|}{TBar}  & \multicolumn{2}{c}{Recoder} \\\cline{2-7} 
        Sub-category& easy & medium & easy & medium & easy & medium                \\\hline\hline
        S-O   & 7 & 5  & 2 & 2  & 2 & 3 \\\hline 
        S-C   & 3 & -  & - & - & 1 & - \\\hline 
        S-V   & 2 & -  & 1 & -  & 1 & - \\\hline
        S-A   & 1 & -  & - & -  & - & - \\\hline
        S-F   & 2 & 1  & - & -  & - & - \\\hline
        S-AS  & 8 & 1  & - & -  & - & 1 \\\hline
        S-DS  & 2 & 3  & 2 & 1  & 2 & 1 \\\hline
        S-HO  & 5 & 5  & 1 & -  & - & - \\\hline
        
\rowcolor{grey}   
        Total  (Single-Hunk) & 30 & 15  & 6 & 3  & 6 & 5  \\\hline

M-S/M-U/M-L   & 31    & 31      & -    & -      & -    & -           \\\hline
\end{tabular}
\vspace{-6pt}
\label{tab:data_apr}
\end{table}





\begin{tcolorbox}[left=2pt,right=2pt,top=2pt,bottom=2pt]
\vspace{-3pt}

Existing pattern based and learning based APR are ineffective at fixing auto-generated code, challenges include: (1) limited search space; (2) unable to generate multi-edit patches; (3) lack of awareness of program dependencies.

\vspace{-3pt}

\end{tcolorbox}


%% file: rq3.tex
\begin{table}[!th]
\centering
\setlength{\tabcolsep}{1.8pt}
\caption{The number of correctly fixed solutions using Codex-e, refer Table \ref{tab:bug-pattern} for abbreviation of defect classification}
\vspace{-6pt}
    \begin{tabular}{c|l|rr|rr|rr|rr}
         \hline
Defect   & Sub          & \multicolumn{2}{c|}{Total} & \multicolumn{2}{c|}{Codex-e$^{bug}$} & \multicolumn{2}{c|}{Codex-e$^{line}$} & \multicolumn{2}{c}{Codex-e$^{stm}$} \\\cline{3-10} 
Category &  -Category   & easy & medium              & easy & medium                        & easy & medium                         & easy & medium \\\hline\hline
\multirow{7}{*}{}
& S-O   & 7    & 5      & 4    & 3      & 1    & 2      & 2    & 4      \\\cline{2-10} 
& S-C   & 3    & -      & -    & -      & 1    & -      & 1    & -      \\\cline{2-10} 
 & S-V  & 2    & -      & -    &  -      & 1    & -      & 1    & -      \\\cline{2-10}
& S-A   & 1    & -      & 1    & -      & -    & -      & -    & -      \\\cline{2-10} 
Single  & S-F   & 2    & 1      & 1    & -      & 2    & -      & 2    & -      \\\cline{2-10} 
-Hunk        & S-AS             & 8    & 1      & -    & -      & -    & -      & 1    & -      \\\cline{2-10} 
             & S-DS & 2    & 3      & 2    & -      & 1    & -      & 2    & -      \\\cline{2-10}
            & S-HO   & 5    & 5      & -    & -      & 1    & -      & 1    & -      \\\hline
\rowcolor{grey}        
Total        & -  & 30   & 15      & 8    & 3      & 7    & 2      & 10   & 4      \\\hline\hline
\multirow{3}{*}{}
Multi        & M-S  & 7    & 2      & 1    & -      & 1    & -      & 2    & -      \\\cline{2-10}
-Hunk        & M-U  & 19    & 20      & 1    & 2      & -    & 1      & -    & -      \\\cline{2-10}
        & M-L    & 5    & 9      & -    & -      & -    & -      & -    & -      \\\hline
\rowcolor{grey}
Total        & -  & 31    & 31      &2    & 2      & 1    & 1      & 2    & -      \\\hline
\end{tabular}
\vspace{-6pt}
\label{tab:data_codexe}
\end{table}

\section{RQ3: Can Codex edit mode fix program bugs?}
\label{sec:rq3}

Recently, OpenAI released a new edit mode of Codex which has the ability to change the content of an existing program. Codex edit mode takes a program and a natural language instruction as inputs, and outputs an edited program based on the instruction. As Codex-e can edit the content of programs, a natural question to ask would be ``Can Codex-e fix an incorrect program with proper instructions?'' 
We designed three strategies to construct the edit instruction for Codex-e. 
\begin{itemize}[leftmargin=*]
    \item \textbf{Codex-e$^{bug}$}: We tell Codex-e that a bug exists in the given program and ask Codex-e to fix it. The instruction is simply given as ``\emph{Fix bug in the program}''.
    \item \textbf{Codex-e$^{line}$}: 
    The instruction for Codex-e is formulated as ``\emph{Fix line N}''.
    \item \textbf{Codex-e$^{stm}$}: 
    We use the program text of the statements, say {\em s1} at the suspicious line and formulate the instruction to Codex-e as ``\emph{Fix s1}''. 
\end{itemize}
For example, to fix the constant mutation bug for  \textit{makeFancyString} in Figure~\ref{listing:tbar_vs_recoder}, we give Codex-e$^{line}$ the instruction \emph{Fix line 6}, and provide Codex-e$^{stm}$ the instruction \emph{Fix ``i -= 2;''}.

For each incorrect solution (we exclude solutions that produce syntax errors as in Section~\ref{sec:rq2}), we select the ten most suspicious statements and ask Codex-e to generate five possible edits for each statement (i.e., Codex-e tries to fix an incorrect solution within 50 attempts). Similar to the initial solution generation in the regular Codex mode, we set the temperature at 0.8 to increase the possibility of finding a correct edit. 

Table~\ref{tab:data_codexe} shows the results for the three strategies, where columns Codex-e$^{bug}$, Codex-e$^{line}$ and Codex-e$^{stm}$ show the number of correct patches using corresponding edit instructions.
With \emph{Fix bug in the program} as instruction, Codex-e$^{bug}$ only learns about the existence of bugs in the given program without any information about the fault locations. Surprisingly, with limited guidance, Codex-e$^{bug}$ successfully produced 15 correct patches where four of these patches involve multi-hunk modifications (refer to supplementary material for the example). 
In contrast, when giving the faulty line number as instruction, Codex-e$^{line}$ fixes nine solutions that require a single-hunk fix, and two solutions that requires a multi-hunk fixes.
Compared to Codex-e$^{bug}$ and Codex-e$^{line}$, Codex-e$^{stm}$ produces the best results by successfully fixing 16 buggy solutions.
We attribute the effectiveness of Codex-e$^{stm}$ to its use of program texts (e.g., ``i -= 2;'') that may guide a language model like Codex in matching relevant statements.
\vspace{-6pt}
\begin{figure}[!h]
\centering
\begin{lstlisting}[style=mystyle, escapechar=@]
public static int[][] construct2DArray(int[] original, int m, int n) {
 // Instruction: Fix "for (int i=0; i<result.length; i++){"
+if (n*m != original.length) @ \color{blue}\ttfamily// S-AS-8@
+  return new int[0][0];
 int[][] result = new int[m][n];
 for (int i=0; i<result.length; i++){
  for (int j=0; j<result[i].length; j++){
-  if (i*result[0].length+j >= original.length)
-   return new int[0][0];
-  else
    result[i][j]=original[i*result[0].length+j];}}
 return result;}
\end{lstlisting}
\vspace{-6pt}
\caption{Flexible fault localization example of LeetCode programming task \textit{convert-1d-array-into-2d-array} on biweekly-contest-62 fixed by Codex-e$^{stm}$}
\label{listing:fixlocation}
\end{figure}

Furthermore, we manually analyze patches produced by Codex-e, and find that Codex-e is able to generate patches at \emph{flexible locations}. Prior APR work \cite{tbar,recoder,lutellier2020coconut,jiang2021cure} have shown a significant performance gap with/without perfect fault localization results. While existing APR tools strictly try to produce patches at a given faulty line number, ignoring the possibility of fixing a bug in the relevant context, Codex-e does not have such limitations. In the 16 correctly fixed solutions by Codex-e$^{stm}$, 8 (50\%) of them are fixed by editing beyond the statement provided in the given instruction. Figure~\ref{listing:fixlocation} shows one such example. The instruction provided to Codex-e$^{stm}$ is \emph{Fix ``for(int i =0; i$<$result.length; i++)\{''}, and Codex-e$^{stm}$ fixes this by moving one \textit{if}-then clause out of the loop body and changing the \textit{if}-condition.
Compared to traditional APR tools, using flexible fault localization is an important feature enabling Codex-e to produce more correct patches.

\begin{tcolorbox}[left=2pt,right=2pt,top=2pt,bottom=2pt]
\vspace{-3pt}
The effectiveness of Codex-e with a given specific fault location (Codex-e$^{stm}$) is nearly comparable to its effectiveness without any location guidance (Codex-e$^{bug}$). 
%

\vspace{-3pt}
\end{tcolorbox}

\begin{figure}[!t]
    \centering
    \includegraphics[width=0.35\textwidth]{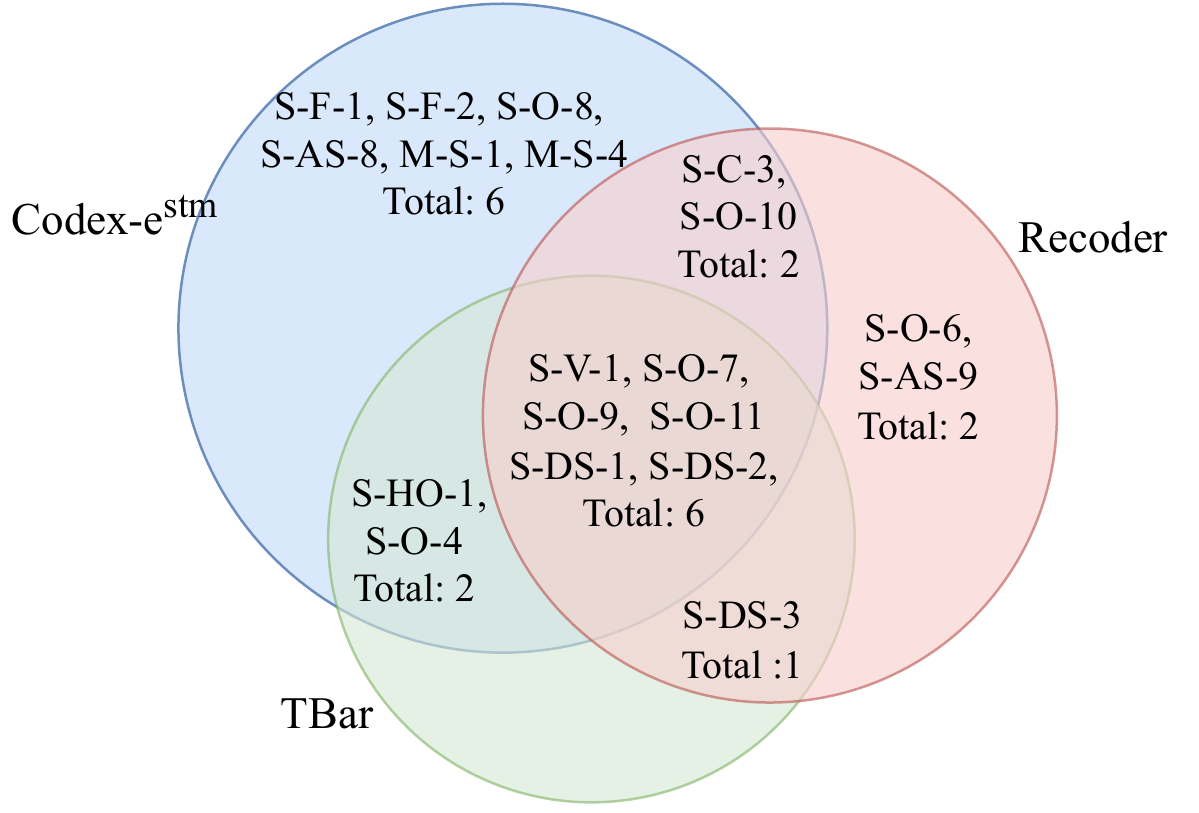}
    \vspace{-6pt}
    \caption{The repair results of different APR tools}
    \label{fig:venn_result}
    \vspace{-6pt}
\end{figure}

\subsection{Comparison between TBar, Recoder and Codex-e$^{stm}$.} To analyze the types of defects fixed by each tool and the reasons behind the effectiveness of each approach, we compare the patches produced by TBar, Recoder, and Codex-e$^{stm}$. As our experiments show that Codex-e$^{stm}$ gives the best overall results among all strategies of Codex-e, we select Codex-e$^{stm}$ for comparison with other APR tools. 
Figure~\ref{fig:venn_result} shows a Venn diagram to better illustrate the set of commonly and uniquely produced patches by these three tools. We denote the set of patches produced by TBar as $TBar$, patches produced by Recoder as $Recoder$, and patches produced by Codex-e$^{stm}$ as $Codex$-e$^{stm}$.
As shown in Figure~\ref{fig:venn_result}, the patches produced by $TBar$ is a proper subset of $Codex$-e$^{stm}$ $\cup$ $Recoder$. 
In fact, the patches produced by $TBar$ is almost subsumed by the set $Recoder$.
This is due to the restricted search space of pattern-based approaches (discussed in Section~\ref{sec:rq2}).
If we compare Codex-e$^{stm}$ and Recoder, both approaches share eight common patches, while Codex-e$^{stm}$ has six more unique patches, and Recoder has two more unique patches. We think that Codex-e$^{stm}$ outperforms Recoder because: (1) Codex-e$^{stm}$ can produce complex patches at flexible locations (e.g., Figure~\ref{listing:fixlocation}); and (2) Codex-e$^{stm}$ is trained on a much larger dataset than Recoder (Recoder uses 82868 human patches for training), which helps Codex-e$^{stm}$ learn more fix patterns (e.g.,  Fig~\ref{listing:codex_unique} where Codex-e$^{stm}$ uses a lambda expression). 

\begin{figure}[!h]
\centering
\begin{lstlisting}[style=mystyle, escapechar=@]
public static int minimumSum(int num){
  // (using lambda expresion) (S-F-1) 
- Collections.sort(digits); 
+ Collections.sort(digits, (a, b) -> b - a); }
\end{lstlisting}
\vspace{-6pt}
\caption{An example that is uniquely fixed by Codex-e$^{stm}$}
\label{listing:codex_unique}
\vspace{-6pt}
\end{figure}


Despite being trained with less data, Recoder still produces two unique patches. Figure~\ref{listing:failtofix4} shows one of the uniquely fixed solutions by Recoder. We think that Recoder can generate this correct fix due to its syntax-guided decoder that can guide it to copy the statement at line 6 and insert it at line 3 of Figure~\ref{listing:failtofix4} (this invokes the copy operation of Recoder that copies the AST subtree rooted at the \texttt{set.remove(i)} statement). In another example (S-O-6) uniquely fixed by Recoder, it correctly replaces a branch condition of the form \texttt{if (a \&\& b)} with \texttt{if (a)} (which is also an AST edit operation). These examples show that encoding AST information into deep learning model may help in generating correct patches. In future, researchers can consider {\em incorporating AST information into large language 
model like Codex-e and AlphaCode}. 





\begin{figure}[!h]
\centering
\begin{lstlisting}[style = mystyle, escapechar=@]
public static List<List<Integer>> findWinners(int[][] matches) {  ...
  for (int i : map.keySet()) {
+   set.remove(i); @\color{blue}\ttfamily// (S-AS-9)@
    if (map.get(i) == 1) {
      ans1.add(i);
      set.remove(i);
      }  }  ...}
\end{lstlisting}
\vspace{-6pt}
\caption{An example that is uniquely fixed by Recoder}
\label{listing:failtofix4}
\vspace{-6pt}
\end{figure}


%% file: rq5.tex
\subsection{Combine Patch Space of Different Tools.}
\label{subsec:combineapr}

\paragraph{\textbf{Combine patch space of Codex-e and APR}}
We further study whether the patch search space produced by APR and Codex-e complement each other by evaluating the patch ingredients produced by different tools.
The \emph{patch ingredient} is defined to be the set of operators/operands (e.g., variables, literals, operators and etc.) used to construct the corresponding patch.
If APR and Codex-e produce patch ingredients that complement each other, their combination will be more likely to generate the correct patch.
To do so, for each incorrect solution, we first obtain the \emph{required patch ingredients} $I_{correct}$ by referring to the ``ground truth'' patch constructed in Section~\ref{sec:rq1} (i.e., the correct patch is built using ingredients in $I_{correct}$).
Then, we investigate the following; we do not consider Recoder+Codex or Recoder+Codex-e since both Recoder and Codex/Codex-e are learning based tools.
\begin{enumerate}[leftmargin=*]
\item Can an individual tool (TBar/Codex-e) produce all required patch ingredients for each incorrect solution?
\item Can combining TBar and Codex-e (run TBar and Codex-e sequentially) produce all required patch ingredients?
\end{enumerate}

Table~\ref{tab:codex_and_apr} shows the number of incorrect solutions whose required patch ingredients are covered by the patch space of each APR technique.
``TBar+Codex-e'' represents the combined patch space of TBar and Codex-e.
Our results show that combining Codex-e and TBar could successfully generate the required patch ingredients of 9 incorrect solutions (with 2 of them cannot be generated by TBar and Codex-e separately).


 
 
 \begin{table}[!t]
\centering
\caption{The number of incorrect solutions that TBar and Codex-e can produce all required patch ingredients, refer Table \ref{tab:bug-pattern} for abbreviation of defect classification}
\vspace{-6pt}
\setlength{\tabcolsep}{2pt}
    \begin{tabular}{l|rrrrr}
         \hline
        Defect  Sub-category & Total & TBar & Codex-e & TBar+Codex-e & TBar+Codex \\\hline\hline
        S-HO & 10 & 1 & 3 & 4 & 5 \\\hline
        M-S  & 9 & 3 & 2 & 4 & 4 \\\hline
        M-U  & 39 & - & - & 1 & 3 \\\hline
\rowcolor{grey}   
      Total  & 58 & 4 & 5 & 9 & 12 \\\hline
\end{tabular}
\label{tab:codex_and_apr}
\vspace{-6pt}
\end{table}

 
Figure~\ref{listing:smallestnumber} shows an incorrect solution that can be fixed by changing $n{>}0$ to $n!{=}0$ \textbf{and} inserting a bound check $\mathit{nums.size}(){>}0$.
For this incorrect solution, none of the APR tools in our experiment generates a correct fix.
However, the two required patch ingredients could be separately produced by TBar and Codex-e. Specifically, TBar fixes the first bug by changing the incorrect operator from ``$>$'' to ``!='' which makes the solution pass the public tests. When we submit this partially fixed solution to~\leet, the program still fails by throwing \texttt{IndexOutOfBoundsException}. 
By encoding the error message into the edit instruction (``Fix IndexOutOfBoundsException''), Codex-e successfully fixes the bug by appending the check \texttt{nums.size()$>$0}.
 
 \begin{figure}[!th]
\centering
\begin{lstlisting}[style=mystyle, escapechar=@]
public static long smallestNumber(long num) {
  long n = num;
  ArrayList<Integer> nums = new ArrayList<>();
- while(n > 0){ @\color{blue}\ttfamily// Fixed by TBar@
+ while(n != 0){ 
    nums.add((int)(n % 10));
    n = n / 10;
  }
  Collections.sort(nums);
- if(nums.get(0) == 0){ @\color{blue}\ttfamily// Fixed by Codex-e@
+ if(nums.size() > 0 && nums.get(0) == 0){
    for(int i = 1; i < nums.size(); i++){ ...
\end{lstlisting}
\vspace{-6pt}
\caption{Combined patch space of TBar and Codex-e}
\label{listing:smallestnumber}
\vspace{-6pt}
\end{figure}

 \paragraph{\textbf{Combine APR with Multiple Solutions of Codex}}
Codex generates a set of program candidates --- each of them may slightly vary in their understanding of the problem description and hence represent slightly different code.
We study the feasibility of combining the patch ingredients from these candidates (``TBar+Codex'' setting). 
Table~\ref{tab:codex_and_apr} shows that ``TBar+Codex'' is the most effective among the evaluated combinations by producing all required patch ingredients for 12 incorrect solutions. 
Figure~\ref{listing:minstep} shows an example incorrect solution which requires two patch ingredients. TBar generates the first patch ingredient (i.e., removing the if-branch from lines 9--11) but the second patch ingredient (adding a for-loop to calculate the sum of absolute value of array \texttt{freq}) does not exist in the patch space of any APR technique (including Codex-e).
This is mainly because generating such a large and unseen code snippet is not supported by most of the existing APR tools, and Codex-e does not have a relevant hint in instruction. 
However, Codex produces many candidate solutions that can be used for enriching the patch space. By borrowing the code from another candidate solution, and modifying the variable name, we can successfully fix the below incorrect solution.
 
\begin{figure}[!h]
\centering
\begin{lstlisting}[style=mystyle, escapechar=@]
public static int minSteps(String s, String t) {
  int[] freq = new int[26];
  for(char c : s.toCharArray())
    freq[c - 'a']++;
  int steps = 0;
  for(char c : t.toCharArray())
-   if(freq[c - 'a'] == 0) @\color{blue}\ttfamily// Fixed by TBar@
-     steps++;
-   else
      freq[c - 'a']--;
// Find in another candidate solutions of minSteps
+ for(int fr : freq)
+   steps += Math.abs(fr);
  return steps;
}
\end{lstlisting}
\vspace{-6pt}
\caption{Obtaining patch ingredients from multiple candidate solutions}
\label{listing:minstep}
\vspace{-6pt}
\end{figure}


Compared to fixing incorrect solutions with only APR techniques, both Codex-e and Codex's multiple solutions could provide required patch ingredients to construct correct fixes.

\begin{tcolorbox}[left=2pt,right=2pt,top=2pt,bottom=2pt]
\vspace{-3pt}
By using patch ingredients extracted from (1) TBar's and Codex-e's patches, and (2) TBar's patches and multiple generated solutions by Codex --- we successfully identify the required patch ingredients of more incorrect solutions.
\vspace{-3pt}
\end{tcolorbox}

%% file: rq4.tex
\section{Implications and Discussions}
Our study identifies several important implications and suggestions for the language models and program repair research.

\subsection{Open dataset for language model defects.} To push the limits of the code generation capability of a large language model like Codex, we believe that our systematic investigation of the mistakes made by language models is an important initial step. It would be beneficial to have a community-driven dataset and more analysis of the defects within the dataset to facilitate future improvement of the auto-generated programs. We propose the \mydataset{} dataset as an initiative towards this direction. 

\subsection{Negative symptoms of auto-generated Codex programs.}  We have identified several negative symptoms among auto-generated Codex programs, including code that contains: (1) names that indicate wrong algorithms, (2) repeatedly producing similar code blocks, (3) irrelevant function helpers. Moreover, we observed that even after manually fixing all auto-generated Codex programs with syntax errors of bracket mismatches, these programs are still incorrect as they fail to pass the held-out tests in \leet{}. 

\subsection{Use of function names in auto-generated code.} 
Based on our manual analysis of the generated solutions, Codex seems to rely heavily on the function name for solving the programming tasks (e.g., \texttt{minimumOperations} in Figure~\ref{listing:example-dp}) . In fact, a recent study has also observed the tendency of Codex in generating solutions based on function name~\cite{jones2022capturing}. Compared to the long prompt (function signature and the problem description), the function name is more concise and easier to search in GitHub. However, this strategy fails when a customized algorithm is required to solve a programming task. Relying on the function's name to search for relevant code will reduce the generation power of Codex to a simple API search engine that returns the implementation for a given API.   \emph{Future language models designed for code generation should focus on summarizing useful information from problem description to reduce reliance on function names.}


\subsection{Pattern-based APR versus learning-based APR.}

Section~\ref{sec:rq2} shows that Recoder generates a few more correct fixes than TBar. The reasons are that pattern-based APR requires (1) additional fix patterns, or (2) a large search space for fix ingredients (e.g., specific literal). Figure~\ref{listing:tbar_vs_recoder} shows an example that can be uniquely fixes by Recoder by adding a statement $\mathit{steps}++;$ at line 19, which is not supported by TBar. However, Figure~\ref{fig:venn_result} shows that TBar also uniquely fixes two solutions where Recoder fails. For example, although Recoder has used the operator mutation for fixing other bugs, it fails to fix incorrect solution (S-O-4) that requires changing the relational operator in ``$\mathit{<}$'' to ``!='' where TBar succeeded. This indicates learning-based APR cannot guarantee a learned pattern is always correctly applied in fixing all programs. \emph{Future APR research on designing fixing operators could work on either (1) incorporating domain-specific knowledge into learning new patterns and (2) improving the generalizability of learned patterns}.

\subsection{How can APR research help language models?} Although our study shows that existing APR techniques can only help to fix a small number of bugs in auto-generated programs by Codex, we believe that APR research can benefit future research in language models in the following aspects: \begin{description}[leftmargin=*]
    \item[\emph{Test-driven repair framework.}] 
   Our study adapts the test-driven repair framework~\cite{qi2015analysis,genprog12} that relies on the quality of test cases, and our results show that the public tests (input/output examples in Figure~\ref{listing:example-dp}) in \leet{} can guide APR tools to generate correct fixes for Codex programs. Specifically, our study shows that we can apply test-driven repair for (1) fixing incorrect solutions generated by the original mode of Codex, and (2) guiding Codex-e by using fault localization information to generate more correct fixes. Language models currently produce a new program from scratch using only natural language instructions. Instead of producing the correct program from scratch, future code generation can first produce an edit of the incorrect program, and further refine it via an iterative test-driven approach.
    \item[\emph{Prioritization of correct programs.}] Our study shows that several negative symptoms exist in auto-generated Codex programs. As our study shows that auto-generated programs with these symptoms are unlikely to lead to correct programs, future designers of language models can integrate a filter function into the language model to automatically eliminate programs with negative symptoms. Another alternative solution is to encode these symptoms into the ranking function to guide the language model in selecting better programs. Both of these directions indicate the potential of incorporating recent advancement of APR research in patch correctness assessment~\cite{anti-pattern,wang2020automated} and patch prioritization~\cite{ye2021automated,ghanbari2020objsim} to guide language models like Codex in generating better programs.
     \item[\emph{Obtaining patch ingredients.}] Our study in Section~\ref{subsec:combineapr} shows that we can effectively combine the patch space of TBar and Codex/Codex-e to obtain the required patch ingredients for generating complex fixes. Future research can work on automatically searching and merging the patch ingredients in generating more complex programs/patches.
\end{description}

\subsection{Balance between control / flexibility for guiding Codex-e.}
Section~\ref{sec:rq3} shows that patches produced by Codex-e rely heavily on the types of provided edit instruction. Compared to Codex-e$^{bug}$ and Codex-e$^{stm}$, Codex-e$^{line}$ generates the least number of correct fixes.  
Although the number of fixed solutions by Codex-e$^{bug}$ and Codex-e$^{stm}$ are quite close (15 versus 16 bugs), the fixed defect category varies. Codex-e$^{bug}$ fixes two more multi-hunk bugs, whereas Codex-e$^{stm}$ fixes three more single-hunk bugs. Since edit instruction like \emph{Fix bug in the program} does not indicate a specific edit target, Codex-e may search for the statements to edit across the entire program based on its learned knowledge. The flexibility encourages generation of large patches but also may lose precision when fixing bugs that require single-line fixes. In contrast, Codex-e$^{stm}$ is provided with a code context (given by fault localization), which steers the edit to the direction that change the most relevant code context. In another perspective, we can also regard Codex-e$^{stm}$ and Codex-e$^{line}$ as test-based APR tools that fix bugs based on fault localization given by test cases, whereas Codex-e$^{bug}$ generates edits without guidance.  Encoding the suspicious code context into the instruction provides more controls and performs better at fixing simple bugs, whereas providing general instruction may find more complex and larger edits due to the increased flexibility. In the future, it is worthwhile to study how to construct edit instructions to guide Codex-e in generating more correct fixes. 



%% file: threats.tex
\section{threats to validity}

\paragraph*{External} During the defect categorization, we eliminate the potential bias by first asking two annotators (two authors of the paper) to manually construct and cross-validate the ``ground truth patch'', if there is any disagreement on patch or defect classification result for a $S_{buggy}$, they further discuss with the other authors to resolve any unclear categorization (e.g., when multiple fixes exist for a bug) until a consensus is reached. We also release our dataset and classification result for public verification. As the performance of the Codex model and repair tools may varies in different settings, our experiments may not generalize beyond the studied configurations and other programming languages beyond Java. We mitigate this threat by reusing configurations given in prior work, and evaluating on several APR tools that use different algorithms (e.g., search-based and learning-based). Although other large language models (e.g., AlphaCode~\cite{alphacode}) exist, our study only evaluates on the Codex language model and the Codex edit mode. The reported findings may not generalize beyond the studied model. As the underlying algorithm used in Codex-e has not been documented, we only use it as a black-box APR tool that produces patches by editing existing programs. To ensure that the training data does not overlap with the evaluated tasks, we have confirmed with the developer of Codex-e that Codex and Codex-e use the same dataset for training. Nevertheless, our experiments show that Codex-e is able to generate fixes for many incorrect solutions. 

\paragraph*{Internal} Our automated scripts may have bugs that can affect our reported results. 
To mitigate this threat, we will make our scripts available upon acceptance. 

%% file: related.tex
\section{Related Work}


\noindent 
\paragraph*{Automated Program Repair}
Automated Program Repair (APR) has gained a lot of attention from both academia and industry in recent years~\cite{cacm19}. APR techniques include search-based, semantic-based and learning-based APR.
Search-based APR tools~\cite{relifix,droix, capgen, simfix, F1X, arja, tbar, liu2019avatar} (e.g., GenProg~\cite{genprog}) take a buggy program and a correct criteria as inputs, and generate patches in two steps: (1) producing patches using predefined code transformation operators; and (2) searching for a patch over the patch space that satisfies a correctness criteria (e.g. passes given tests).
Search-based repair can scale to large programs, but often not to large search spaces. Semantics-based APR techniques (e.g.,
SemFix~\cite{semfix}, Nopol~\cite{NOPOL}, and Angelix~\cite{angelix}) generate patches by (1) formulating a repair constraint that needs to be satisfied by a program passing a given test-suite; and (2) solving the repair constraint to generate patches.
The application of deep learning techniques in program repair has been explored in past few years. DeepRepair~\cite{deeprepair} and DeepFix~\cite{gupta2017deepfix} are the early attempts to fix bugs by learning fixes from similar code. SequenceR~\cite{chen2019sequencer} adapts neural machine translation (NMT) to generate patch, whereas CoCoNuT~\cite{lutellier2020coconut} and CURE~\cite{jiang2021cure} further improve the results by either encoding program context or using a programming language model. DLFix~\cite{li2020dlfix} uses two-layer tree-based RNN to learn code transformations, and Recoder~\cite{recoder} designed a syntax-guided learning approach to improve the decoder of a DL model. In this work, we select Recoder because it fixes the most number of bugs in Defects4J~\cite{defects4j} among those DL-based APR tools whose training model is publicly available.

\paragraph*{Large Language Model for Code Generation} Large language models such as GPT-3~\cite{gpt3} have shown promising performance in the NLP domain. Hendrycks et al.~\cite{apps} proposed APPS dataset and evaluated the code generation performance of several variant GPT models with APPS as the fine-tuned data. 
Later Codex \cite{codex}, the back-end model that powers GitHub Copilot, Alphacode \cite{alphacode}, Codewhisperer \cite{codew}, and \cite{googlellmsynthesis} have emerged as language model based automatic code generation platforms.
There are emerging approaches combining program synthesis with large language model on fixing API usage~\cite{jigsaw} and synthesizing regular expression~\cite{rahmani2021multi}, whereas we focus on fixing general errors in code from programming competitions. Nguyen et al.~\cite{nguyen2022empirical} evaluated the quality of code generated by Copilot on a small set of randomly selected \leet~programming tasks (33 tasks with 132 solutions). Compared to their work, we performed a detailed analysis of $113$ programming tasks via the larger dataset~\mydataset{} which we built. The most relevant papers to us are studies on how language model can fix bugs~\cite{codex_security, codex_apr};  we evaluated whether APR tools (including and combining Codex-e) can fix programs automatically produced by Codex.


%% file: conclusion.tex
\section{Perspective}

In this paper, we study the mistakes made by auto-generated programs from language models like Codex, and  investigate whether automated program repair (APR) tools  can fix the auto-generated buggy programs.
Our study of code generated from language models reveal that: (1) programs produced by Codex share common defect categories as human programmers; (2) existing APR tools (TBar and Recoder) do not perform well at fixing bugs in auto-generated programs
(3)  given proper instructions such as information from fault localization, Codex edit mode (Codex-e) shows promising results in code edit generation, which outperforms TBar and Recoder.
Our study leads us to the following view-points:
\begin{itemize}
    \item 
We suggest enhancing language models with software engineering artifacts such as fault location, with the goal of generating higher quality code.
\item We suggest directions for automated program repair (APR) research inspired by language models, such as 
(i) extracting patch ingredients from automatically generated solution set of Codex, and (ii) making fault localization (fix localization) in program repair more flexible.
\end{itemize}


%% file: exapme-table2.tex
\section{Examples for Table-2}

\subsection{Multi-Hunk Similar (M-S)}

\begin{figure}[!h]
\centering
\begin{lstlisting}[style = mystyle, escapechar=@]
    public static List<List<Integer>> findDifference(int[] nums1, int[] nums2) {
        /** Given two 0-indexed integer arrays nums1 and nums2, return a list answer of size 2 where:
         *  answer[0] is a list of all distinct integers in nums1 which are not present in nums2.
         *  answer[1] is a list of all distinct integers in nums2 which are not present in nums1.
         *  Note that the integers in the lists may be returned in any order.
         */
        List<List<Integer>> ans = new ArrayList<>();
        Set<Integer> set1 = new HashSet<>();
        Set<Integer> set2 = new HashSet<>();
        for (int n : nums1) {
            set1.add(n);
        }
        for (int n : nums2) {
            set2.add(n);
        }
        for (int n : nums1) {
            if (set2.contains(n)) {
                set2.remove(n);
            }
        }
        for (int n : nums2) {
            if (set1.contains(n)) {
                set1.remove(n);
            }
        }
        ans.add(new ArrayList<>(set1));
        ans.add(new ArrayList<>(set2));
        return ans;
    }
\end{lstlisting}
\vspace{-6pt}

\caption{Examples of Multi-Hunk Similar}
\label{listing:multi-hunk-similar}
\vspace{-6pt}
\end{figure}

\subsection{Multi-Hunk Unique (M-U)}

\begin{figure}[!h]
\centering
\begin{lstlisting}[style = mystyle, escapechar=@]

\end{lstlisting}
\vspace{-6pt}

\caption{Examples of Multi-Hunk Unique}
\label{listing:multi-hunk-unique}
\vspace{-6pt}
\end{figure}

\subsection{Multi-Hunk Large Fix (M-L)}

\begin{figure}[!h]
\centering
\begin{lstlisting}[style = mystyle, escapechar=@]

\end{lstlisting}
\vspace{-6pt}

\caption{Examples of Multi-Hunk Large Fix}
\label{listing:multi-hunk-large}
\vspace{-6pt}
\end{figure}

\subsection{Single-Hunk Operator Mutation (S-O)}

\begin{figure}[!h]
\centering
\begin{lstlisting}[style = mystyle, escapechar=@]

\end{lstlisting}
\vspace{-6pt}

\caption{Examples of Single-Hunk Operator Mutation}
\label{listing:single-hunk-operator}
\vspace{-6pt}
\end{figure}

\subsection{Incomplete Code}

\begin{figure}[!h]
\centering
\begin{lstlisting}[style = mystyle, escapechar=@]

\end{lstlisting}
\vspace{-6pt}

\caption{Examples of Incomplete Code}
\label{listing:incomplete_code}
\vspace{-6pt}
\end{figure}

\newpage

\subsection{Invoke Undefined Program Elements}

\begin{figure}[!h]
\centering
\begin{lstlisting}[style = mystyle, escapechar=@]

\end{lstlisting}
\vspace{-6pt}

\caption{Examples of Invoking an undefined function}
\label{listing:undefined_function}
\vspace{-6pt}
\end{figure}

\subsection{Incompatible types}

\begin{figure}[!h]
\centering
\begin{lstlisting}[style = mystyle, escapechar=@]

\end{lstlisting}
\vspace{-6pt}

\caption{Examples of Incompatible types}
\label{listing:incompatible_type}
\vspace{-6pt}
\end{figure}